\def\etal{{\it et al.\thinspace}}
\def\eg{{\it e.g.\ }}

\def\ie{{\it i.e.\ }}

\def\gsim{~\rlap{$>$}{\lower 1.0ex\hbox{$\sim$}}}
\def\lsim{~\rlap{$<$}{\lower 1.0ex\hbox{$\sim$}}}
\def\mearth{{\rm\,M_\oplus}}

\documentclass[12pt,preprint]{aastex}

\begin{document}

\title{The HD 40307 Planetary System: Super-Earths or Mini-Neptunes?} 

\author{Rory Barnes\altaffilmark{1}, Brian Jackson\altaffilmark{1}, Sean N. Raymond\altaffilmark{2,3}, Andrew A. West\altaffilmark{4}, Richard Greenberg\altaffilmark{1}}

\altaffiltext{1}{Lunar and Planetary Laboratory, University of Arizona,
Tucson, AZ 85721}
\altaffiltext{2}{NASA Postdoctoral Program Fellow}
\altaffiltext{3}{Center for Astrophysics and Space Astronomy, University of Colorado, Boulder,
CO 80309}
\altaffiltext{4}{Astronomy Department, University of California, 601 Campbell Hall, Berkeley, CA 94720-3411}

\keywords{}

\begin{abstract}
Three planets with minimum masses less than 10 $\mearth$ orbit the
star HD 40307, suggesting these planets may be rocky. However, with
only radial velocity data, it is impossible to determine if these
planets are rocky or gaseous. Here we exploit various dynamical
features of the system in order to assess the physical properties of
the planets.  Observations allow for circular orbits, but a numerical integration shows that the eccentricities must be at least
$10^{-4}$. Also, planets b and c are so close to the star that tidal effects
are significant. If planet b has tidal parameters similar to the
terrestrial planets in the Solar System and a remnant eccentricity larger
than $10^{-3}$, then, going back in time, the system would have been
unstable within the lifetime of the star (which we estimate to be
6.1$\pm$1.6 Gyr). Moreover, if the eccentricities are that large and
the inner planet is rocky, then its tidal heating may be an order of
magnitude greater than extremely volcanic Io, on a per unit surface area
basis. If planet b is not terrestrial, \eg Neptune-like, these
physical constraints would not apply. This analysis suggests the
planets are not terrestrial-like, and are more like our giant
planets. In either case, we find that the planets probably formed at
larger radii and migrated early-on (via disk interactions) into their current orbits. This study
demonstrates how the orbital and dynamical properties of exoplanet
systems may be used to constrain the planets' physical properties.
\end{abstract}

\section{Introduction}
Mayor \etal (2008) recently announced the discovery of a system of
three planets, b, c, and d, orbiting the K dwarf HD 40307. This system
is unique because it is the first detected system in which all three
companions have minimum masses less than 10 $\mearth$. Moreover, the
innermost planet, b at 4.2 $\mearth$, has the lowest minimum mass yet detected by radial velocity methods. In addition to relatively low masses, the system is
striking in that the planets appear to lie close to a Laplace-like
resonance: very small eccentricities ($\lsim 0.01$) and period ratios
near 4:2:1. However, as noted by Mayor \etal (2008), the observations
rule out such a resonance chain with high confidence ($>10\sigma$).

If, as is most probable, the actual masses are similar to the minimum value,
it is natural to wonder whether these bodies are larger versions of
the rocky planets in the Solar System (``super-Earths''), or smaller
versions of our gaseous planets (``mini-Neptunes''). The only
currently available method to make a direct assessment of these two
possibilities requires transit data, but none has been reported, so we
must rely on indirect means.  One possibility is to consider the
theoretical modeling of terrestrial and gaseous planet
formation. However, that approach leads to uncertainty. For example,
core-accretion models predict that a solid core requires 2 -- 10
$\mearth$ in order to accrete a hydrogen envelope (Pollack \etal 1996;
Bodenheimer, Hubickyj, \& Lissauer 2000; Ikoma \etal 2001; Hubickyj,
Bodenheimer \& Lissauer 2005). Therefore, planet formation models
cannot yet constrain the physical nature of the $4 - 10 \mearth$
planets in this system.

Here we exploit another method for constraining the properties of
exoplanets: The orbital history since formation. In our Solar System,
the rocky and gaseous planets' responses to tides are very different;
solid, rocky bodies dissipate tidal energy more effectively (smaller
$Q'$ values) than their gaseous counterparts. Tides result in orbital
migration at rates that can be orders of magnitude different depending
on whether the planet is rocky or gaseous (Jackson \etal 2008a,
2008b). Also, the effectiveness of tides falls off rapidly with
distance such that in multiple planet systems with close-in planets,
the innermost planet has usually experienced tidal migration, while
the others have undergone little. By modeling the past tidal evolution
of a system, the inner planets' properties may be constrained by
forbidding past events, \eg mean motion resonance crossing or crossing
orbits, that would have led to orbits inconsistent with the current
system. For example, Barnes \etal (2008) considered the GJ 581 system
(Udry \etal 2007) and showed that planet c ($\gsim 5 \mearth$) could
be terrestrial, but cannot have the same tidal parameters as the
present-day Earth. Jackson \etal (2008b) considered the GJ 876 system
(Rivera \etal 2005) and showed that planet d ($\gsim 7.5 \mearth$)
cannot be terrestrial because, considering tides, $e$
would have been $\approx 1$ less than $\sim 30$ Myr ago, with internal heating
rates up to $10^5$ times that of Io.

For the HD 40307 system, we can exploit the proximity of
mean motion resonances to constrain the tidal evolution of the
innermost planet, b. We focus on two end-member cases for HD 40307 b:
rocky or gaseous, but allow for other possibilities. To exploit
dynamical constraints, we also use what we know about the system's
age, as well as how it may have formed. As we show in the following
sections, the possibility that the planets are terrestrial in
character (\ie ``super-Earths'') seems unlikely, but cannot be ruled
out. However, the gaseous case (\ie ``mini-Neptunes'') is less
constrained. Nor can we rule out an intermediate case, that is tidal
parameters in between those of the terrestrial and gaseous planets in
our Solar System. In $\S$ 2 we describe our dynamical and tidal
models. In $\S$ 3 we present our results for this system. In $\S$ 4 we
infer the character of these planets, and identify likely formation
scenarios. In the Appendix we estimate the age of the star, and, by
extension, the system.

\section{Methods}
\subsection{Planet-Planet Interactions}
We will consider the oscillations of the planets' orbits with the
N-body code \texttt{HNBody},\footnote{Publicly available at
http://www.astro.umd.edu/$\sim$rauch/HNBody.}, which includes general
relativistic effects. For these integrations, we require
numerically-induced energy changes to be less than one part in $10^4$,
which is adequate precision to produce reliable results (Barnes \&
Quinn 2004).

\subsection{Tidal Evolution Models}
For our tidal model we use conventional equations assembled by Goldreich \& Soter
(1966; see also Jackson \etal 2008a; Barnes \etal 2008;
Ferraz-Mello \etal 2008). The evolution of semi-major axis $a$ and
eccentricity $e$ (to second order in $e$) can be modeled as
\begin{equation}
\label{eq:adot}
\frac{da}{dt} = -\Big(\frac{63}{2}\frac{\sqrt{GM_*^3}R_p^5}{m_pQ'_p}e^2 + \frac{9}{2}\frac{\sqrt{G/M_*}R_*^5m_p}{Q'_*}\Big)a^{-11/2}
\end{equation}
\begin{equation}
\label{eq:edot}
\frac{de}{dt} = -\Big(\frac{63}{4}\frac{\sqrt{GM_*^3}R_p^5}{m_pQ'_p} + \frac{225}{16}\frac{\sqrt{G/M_*}R_*^5m_p}{Q'_*}\Big)a^{-13/2}e,
\end{equation}
where $G$ is Newton's gravitational constant, $M_*$ is the stellar
mass, $R_p$ is the radius of the planet, $m_p$ is the planet mass,
$Q'_p$ and $Q'_*$ are the planet's and star's tidal dissipation function
divided by two-thirds its Love number, and $R_*$ is the stellar
radius. In Eqs. (\ref{eq:adot} -- \ref{eq:edot}) the first terms
represent the tide raised on the planet by the star, and the second terms
the tide raised on the star by the planet. This model assumes the
tidal components maintain a constant phase lag from
the line connecting the centers of mass of the two bodies, and is consistent with observations of the
Galilean satellites of our Solar System (Aksnes \& Franklin 2001).  See Jackson \etal (2008a) or Barnes \etal
(2008) for more discussion of this model, but note that other
plausible models, also consistent with observations in the Solar System, exist  (see \eg Hut 1981; N\'eron de Surgy \& Laskar
1997; Eggleton \etal 1998; Mardling \& Lin 2002; Efroimsky \& Lainey
2007; Dobrovolskis 2007; Ferraz-Mello \etal 2008).

Our model does not include the effects of
secular interactions between the planets (see \eg Mardling 2007). Such
effects could potentially play a role, but we will show that our
conclusions regarding the HD 40307 b system are probably not impacted
by the neglect of this effect. If it were included, then the timescale
for eccentricity evolution would likely be slower, and the tidal
evolution of the innermost planet would change the eccentricities of
other planets in the system. (Exterior planets modify the eccentricity
of the inner planet, as tidal evolution damps
it.) Eqs.\ (\ref{eq:adot}) and (\ref{eq:edot}) allow a reasonable description of the tidal evolution of the system.

We may also determine the amount of heat generated in a body due
to tidal friction:
\begin{equation}
\label{eq:heat}
H = \frac{63}{4}\frac{(GM_*)^{3/2}M_*R_p^5}{Q'_p}a^{-15/2}e^2
\end{equation}
(Jackson \etal 2008b). $H$ represents the internal heating rate,
but for geophysical considerations it is useful to express the heat as a surface flux
$h = H/4\pi^2R_p^2$. For reference the heat flux on the Earth (due to radiogenic processes) is 0.08
W/m$^2$ (Davies 1999), Io's is 2 -- 3 W/m$^2$ (Peale \etal 1979;
McEwen \etal 2004), and Europa's, scaling from Io's, could be $\sim 0.2$
W/m$^2$ (O'Brien \etal 2002).

A key parameter is $Q'_p$, which parameterizes the planet's tidal
response to the star. In principle $Q'_p$ may take any value, but in
our Solar System rocky and gaseous bodies tend to cluster around two
$Q'_p$ values separated by several orders of magnitude. For
terrestrial bodies $Q'_p \sim 500$ (Dickey \etal 1994; Mardling \& Lin
2004; Lambeck 1977; Yoder 1995). For gaseous bodies it is common to
adopt $Q'_p \sim 10^5$ (Banfield \& Murray 1992; Aksnes \& Franklin
2001; Zhang \& Hamilton 2007, 2008), although it could be much larger (Greenberg \etal 2008).
The stellar value $Q_*'$ is not very important in the case of HD 40307
because the planet masses and radii are relatively small (see Eqs.\
[\ref{eq:adot} -- \ref{eq:edot}]); we assume it is $3 \times 10^6$
(Jackson \etal 2008a).

We must also estimate stellar and planetary radii. We assume that the
star's radius follows the empirical relationship found by Gorda \&
Svechnikov (1999). For terrestrial cases we scale $R_p$ as
$m_p^{0.27}$ (Fortney \etal 2007). For gaseous cases, $R_p$ is
calculated by assuming the planet has the same mean density as
Neptune: $r_b = (m_b/M_{Nep})^{1/3}R_{Nep}$, where $M_{Nep}$ and
$R_{Nep}$ are the mass and radius of Neptune, respectively. Note that
this assumption is consistent with observations of the transiting
planet GJ 436 b (Deming \etal 2007; Gillon \etal 2007; Jackson \etal
2008a). Furthermore, our analysis depends on the age of this system,
which we estimate as $6.1\pm1.6$ Gyr old (see Appendix).

\section{Results}
Table 1 lists a set of values for the masses and orbits of the planets
in HD 40307 computed by Mayor \etal (2008) with all eccentricities
treated as free parameters. $P$ is the orbital period and $T_p$ is the
time of periastron passage. As $e$ values $\lsim 10^{-2}$ are not
currently measurable in radial velocity data (the uncertainty is
larger than the nominal values), Mayor \etal (2008) prefer a solution
in which all eccentricities are set to zero (and longitudes of
periastron $\varpi$ are therefore undefined). In this case the
residuals dropped slightly, but the other parameters remained the same
(with $T_p$ now the time of passage through longitude zero). Here, we
consider two possibilities for this system, one in which the
eccentricities have the reported non-zero values listed in Table 1 and
one in which the eccentricities are all zero.

\begin{figure}
\plotone{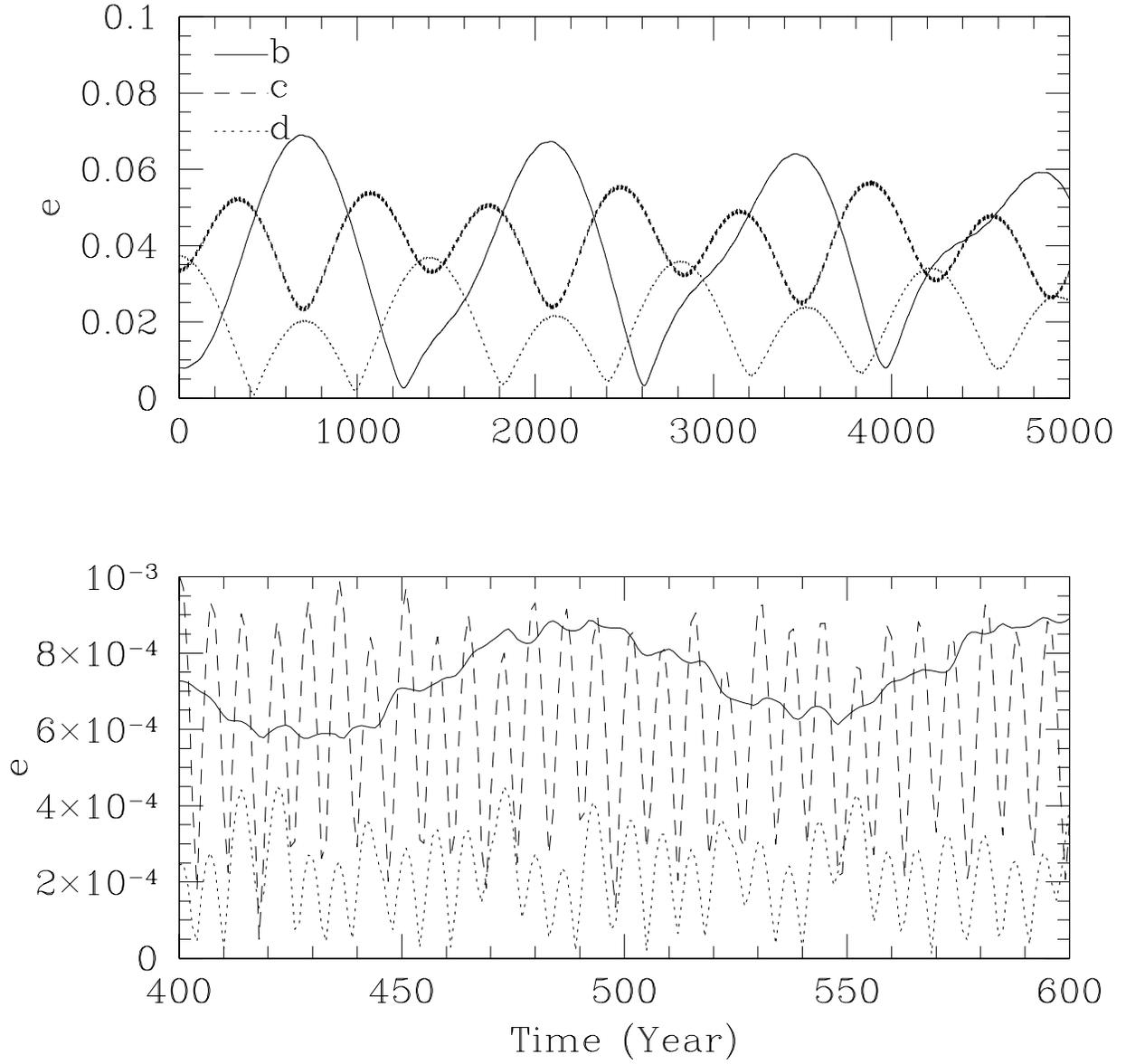}
\figcaption[]{\label{fig:ecc} \small{\textit{Top:} Eccentricity evolution of HD 40307 b, c, and d assuming nominal parameters from Table 1. \textit{Bottom:} Eccentricity evolution assuming all $e$ values are initially zero (the timescale here is shorter than above because the oscillation period is much shorter and was chosen to illustrate the maximum values of $e_b$).}}
\end{figure}

Over timescales much shorter than tidal evolution, interactions among
the planets cause periodic variations in orbital elements. Fig.\
\ref{fig:ecc} shows the variations of the eccentricities, produced by
our N-body models. In the top panel we model the oscillation of the
orbits using the reported, non-zero eccentricities. In the bottom
panel, we show the evolution assuming all eccentricities are initially
zero. For these cases we set $\varpi_b = \varpi_d = 0$ and $\varpi_c
=180^\circ$. (Different choices of the $\varpi$'s can result in about
a factor of two difference in the eccentricity oscillation amplitudes,
but this difference does not affect our conclusions.)  If we assume
the eccentricities are initially nonzero, then they remain below 0.1,
with planet b's eccentricity as large as 0.07. If all the
eccentricities are initially zero, then they all quickly grow to
nonzero values. Even if eccentricities were fully damped by tides,
mutual interactions would keep them $> 0$. The minimum perturbed
eccentricity is thus $\sim 5 \times 10^{-4}$.

\begin{figure}
\plotone{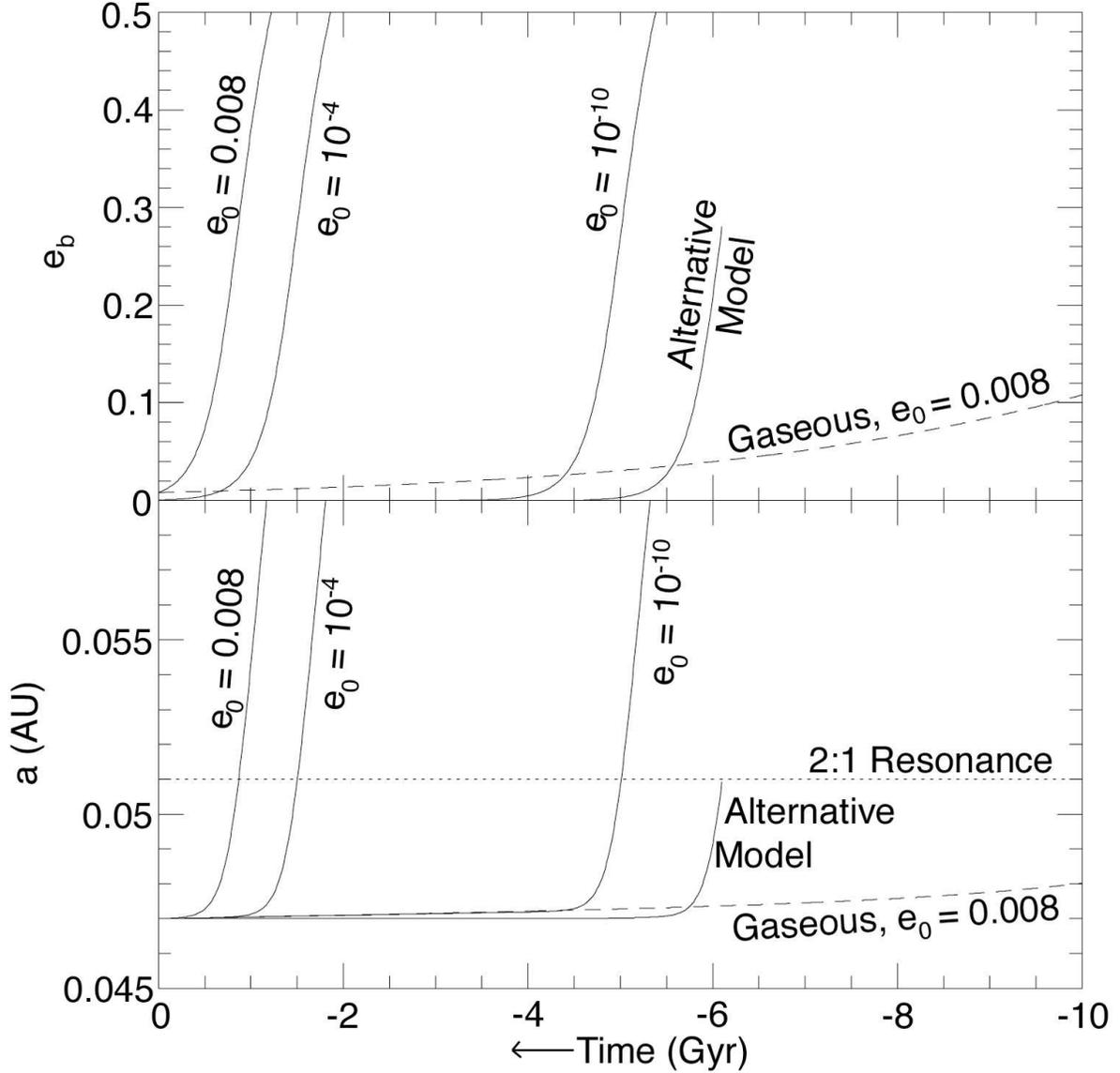}
\figcaption[]{\label{fig:tide} \small{\textit{Top:} History of $e_b$
for different orbital and physical assumptions. Solid curves assume
$Q'_b = 500$ (rocky super-Earths), dashed curves assume $Q'_b = 10^5$
(gaseous mini-Neptunes). The assumed value of the initial eccentricity of planet b, $e_0$, is indicated on each curve. The ``Alternative Model'' curve assumes that planet b formed
with $e \approx 0.25$ and $a_b$ just interior to the 2:1 resonance with
planet c. The value of $e_b$ in this case damped to its minimum perturbed value
and remained there.  \textit{Bottom:} Evolution of $a_b$ (solid and
dashed lines) for the same cases and in the same sequence as
above. For reference the location of the inner 2:1 resonance with
planet c (labeled ``2:1 Resonance'') is shown by the dotted line. Note
that we assume that any history prior to the 2:1 crossing is unphysical.}}
\end{figure}

We have modeled the long-term effect of tides on planet b by
integrating Eqs.\ (\ref{eq:adot} -- \ref{eq:edot}) back in time for
various assumed values of its current eccentricity ($e_0$) (Fig.\
\ref{fig:tide}). (The ``Alternative Model'' is explained below.) For
these three cases, $a_b$ jumps up in values at about the same time as
$e_b$ gets large. So, for example, with $e_0 = 0.008$, $a_b$ would
have been at the location of the 2:1 resonance with planet c less than
1 Gyr ago (dotted line in the bottom panel of Fig.\
\ref{fig:tide}). If planet b crossed the resonance, both planets'
eccentricities would have been pumped up because it is a divergent
crossing (Hamilton 1994; Chiang, Fischer \& Thommes 2002; Zhang \&
Hamilton 2007, 2008). Such a crossing would likely have destabilized
the system, or at least prevented the system from appearing as packed
as it is today. This process would have been similar to models of the
2:1 resonance crossing of Jupiter and Saturn in our Solar System
(Gomes \etal 2005), which significantly spread out the giant
planets. For HD 40307, such a history is unlikely given the current
orbital architecture. Thus, if we assume that the resonance crossing
could not have happened during this system's history, either (i) tides
must have damped $e_0$ to its minimum possible value several billion
years ago, (ii) the system must be younger than 1 Gyr, or (iii) $Q'_b$
must be larger than 500. 

While using $e_0 = 5 \times 10^{-4}$ as an initial condition for the tidal evolution gives resonance crossing 2 Gyr ago, a different history is possible.
The current value of $e$ for this planet would be $\sim 5 \times 10^{-4}$ even if the system
was fully-damped by tides much earlier. Thus, we have no way to
constrain how long the system has been in this state; the $e_0 = 5
\times 10^{-4}$ curve in Fig.\ \ref{fig:tide} could be shifted by any amount to the right and
still be a possible model of the system's history. For example, the
curve labeled ``Alternative Model'' has been shifted so that the system formed inside the 2:1
resonance 6.1 Gyr ago (its estimated age; see Appendix), then was
fully damped within 2 Gyr, and remained with the minimum perturbed
eccentricities ever since (with such small eccentricities, $da/dt \approx 0$ for the intervening 4
Gyr). This scenario avoids a past
resonance and permits super-Earths ($Q'_p \sim 500$) in the HD 40307
planetary system.

Next we calculate the tidal heating of the planets, still assuming terrestrial planet parameters ($Q'_p = 500$, density of 5 g/cm$^3$). For eccentricities of $5 \times 10^{-4}$, the heating is
$4 \times 10^{-3}$ W/m$^2$ for planet b and $10^{-5}$ W/m$^2$ for planet c, assuming radii of $1.07 \ times 10^4$ km and $9.4 \times 10^3$ km, respectively. On the per-unit-surface-area basis (relevant for surface characteristics), these values are considerably lower than the Earth's
radiogenic heating, 0.08 W/m$^2$.  If the planets have the non-zero eccentricities listed in Table 1,
the current surface heat fluxes for these two planets would be $h_b = 1$
W/m$^2$ and $h_c = 0.4$ W/m$^2$, values that lie between the Earth ($\sim$ 0.08 W/m$^2$) and
Io ($\sim$ 2 W/m$^2$). (Note that the $Q'$ value for Io is probably similar to the Earth's [Yoder
1995; Aksnes \& Franklin 2001]). Even though $h_b$ is less than the
heating on Io, it is probably close enough to result in intense
volcanism.

\begin{figure}
\plotone{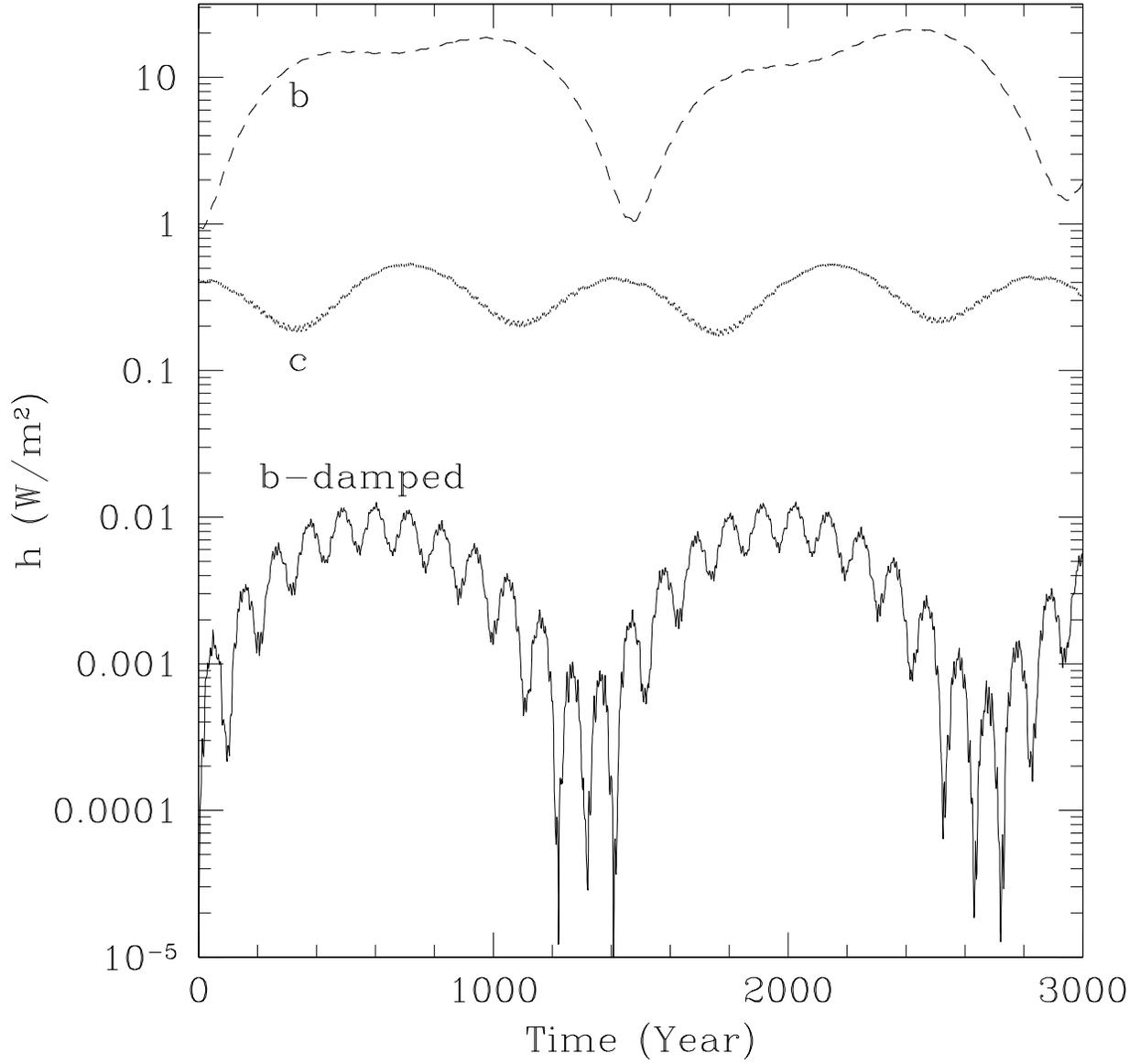}
\figcaption[]{\label{fig:heat} \small{Tidal heating fluxes of terrestrial-like planets in HD 40307. The upper two curves correspond to the changes in $e$ shown in Fig.\ \ref{fig:ecc} (top), which used initial conditions from the nominal orbital elements in Table 1. The curve labeled ``b-damped'' corresponds to the changes in $e$ shown in Fig.\ \ref{fig:ecc} (bottom), in which orbits are initially circular.}}
\end{figure}

The short-term periodic variations of the heating due to the periodic
variations in eccentricities (from Fig.\ \ref{fig:ecc}) are also
interesting. The top two curves in Fig.\ \ref{fig:heat} show the
changes in $h$ that correspond to the orbital changes in the top panel
of Fig.\ \ref{fig:ecc}: $h_c$ oscillates between 0.2 and 0.5
W/$m^2$, always more than the Earth's heating. However, $h_b$ reaches
well over 10 W$/m^2$, and maintains that rate most of the
time. Although the heating rates oscillate, the periods of oscillation
are much shorter than heat transport (predominantly mantle overturn)
in the Earth (see \eg Davies 1999), so the planet's surface flux would
probably be the average heating rates ($\approx 10$ W/m$^2$), still several
times that of super-volcanic Io.

Assuming the system has damped to the minimum eccentricities allowed
by mutual perturbations (Fig.\ \ref{fig:ecc} bottom), the heating
rates, are much lower; $h_b$ is shown in the lowest curve
in Fig.\ \ref{fig:heat}. Here the heating rate oscillates by many
orders of magnitude, but remains much less than the Earth's. Planet
c's tidal heating in this damped case is always $< 10^{-4}$
W/m$^2$. Therefore, if the eccentricities have damped to minimum
values, these planets would have heat fluxes much less than the
terrestrial planets in our Solar System.

For the planets orbiting HD 40307 to be rocky, they must have begun tidal evolution with low eccentricity orbits and with planet b interior to the 2:1 resonance with b. However, it is unlikely that the planets acquired most of their mass in such a configuration, as that scenario predicts an implausibly large pre-planetary nebula. According to Kuchner (2004), such in situ formation requires a primordial disk with
surface density profile $\Sigma (r) = 6379 (r/0.1 {\rm AU})^{-0.925}
\,\, $g\,cm$^{-2}$.  This disk contains 21.6 $\mearth$ inside 0.15 AU, probably
15-100 times more than our Solar System had (Weidenschilling 1977;
Hayashi 1981).\footnote{We assume that $\Sigma = \Sigma_0 (r/1 {\rm
AU})^{-x}$, where ($\Sigma_0,x$) = (7.75 $g \, cm^{-2}$, 1.5) or
(5.895 $g \, cm^{-2}$, 1.0) and extend the disks in to $r=0$.  These
disks are calibrated to the MMSN and contain 5 $\mearth$ from 0.4-4
AU.}  This result is consistent with the assumption that HD 40307's
gas-to-dust ratio is half the solar nebula's because [Fe/H] = -0.31,
in which case its disk would have been 30-200 times more massive than
the solar nebula.  These values represent a disk mass comparable to
the stellar mass, and far in excess of the typical star-disk mass ratio of
1\% (Andrews \& Williams 2005). We conclude that these planets did not
form in situ; they must have formed further out and migrated in prior
to the dispersal of the gas disk.

So far we have considered the implications of rocky planets. Could
they instead be mini-Neptunes with a thick gaseous envelopes and $Q' =
10^5$? If the current $e_b$ has the value given in Table 1, the tidal
history (shown by dashed lines in Fig.\ \ref{fig:tide}) would not have
included dangerous resonance crossings in the last 6.1
Gyr. Furthermore, such planets would have heating fluxes several
orders of magnitude smaller than terrestrial planets, and their
internal structures would probably not be significantly affected by
tidal heat. If, however, the mass (and hence radius) of planet b is
significantly greater than the observational minimum, then there may
have been more evolution. We find that if $m_b = 15 \mearth$, then HD
40307 b would have been at the 2:1 resonance 6.1 Gyr ago. Such a mass
corresponds to an orbit inclined by $75^\circ$ to the line of
sight. In other words, planet b's orbit must be more than 15$^\circ$
from face-on.

If the planets are gaseous, did they form in situ, or did they migrate
in from further out?  If the planets formed in situ, the pre-planetary
nebula would have required about half as much mass as for the
rocky-planet case described above (assuming the cores' masses are
roughly equal to the envelopes' masses), but that value is still
improbably large. More likely, as with rocky planets, the planets
could have formed further out and migrated in via interactions with
the disk (Lin \etal 1996). Theoretical models of this phenomenon (\eg
Snellgrove \etal 2001; Lee \& Peale 2002) suggest resonance capture
could be an outcome of this process, but resonances are not observed
(Mayor et al. 2008). However, such a commensurability could have been
destroyed by subsequent mergers, scattering or turbulence (Terquem \&
Papaloizou 2007; Adams, Laughlin \& Bloch 2008; Lecounet \etal
2008). Therefore the migration scenario is consistent with the
observed orbits. We conclude that if the planets are mini-Neptunes,
they likely formed at larger radii and migrated in via disk torques to
their present orbits.

The terrestrial and gaseous planet models considered above are
not a complete exploration of parameter space. For example, the
responses of the planets to tides are encapsulated in $Q'$, a
notoriously uncertain parameter, even in the Solar System. To address
this uncertainty, we  solved Eqs.\
(\ref{eq:adot} -- \ref{eq:edot}) for a range of values of $Q'_p$ and $e$ ($1500 \le Q'_b
\le 3500$ and $10^{-4} \le e_0 \le 0.01$) to determine how long ago
the resonance crossing would have occurred. If it occurred more than
6.1 Gyr ago, the case is allowed, but if the crossing occurred within
the last 6.1 Gyr, then the case is forbidden. For $e_b < 10^{-3}$,
these restrictions do not strictly apply since the eccentricity may have
damped to those values an arbitrarily long time ago. If, however,
$e_b$ is found to be greater than $10^{-3}$, then it would constrain $Q'_b$ to be
$\gsim 2200$.

\section{Conclusions}
The small minimum masses of the planets orbiting HD 40307 are
tantalizingly close to masses of the rocky, ``terrestrial'' planets in
our Solar System. By considering the dynamical features and history of
the system, we have determined implications of their being
predominantly rocky or gaseous. We find both possibilities are
consistent with the observations, but the likelihood that they are
terrestrial depends on the actual values of the current
eccentricities, which are at or below the detection threshold.

Mayor \etal (2008) report two sets of $e$ values, either all $\sim 0.01$ or all zero. The latter case is ruled out by mutual
perturbations between the planets (see Fig.\
\ref{fig:ecc}). Instead, the lowest possible eccentricities are $\sim
10^{-3}$. If the values are that small, tides probably damped them
from higher values, and they may have reached this state recently, or at
any time in the past. These ambiguities preclude a definitive
assessment of the composition of the planets. However, if or when the
eccentricities are measured more precisely, such a determination will
be possible through tidal analysis.

If the planets are terrestrial (\ie $Q' \sim 500$), then either (a)
the system must be less than 2 Gyr old, which is unlikely (see Appendix), or (b) tidal evolution began with
planet b just inside the 2:1 resonance with c, and with modest
eccentricity ($< 0.3$). In either case, the planets must have formed
at larger distances and migrated inward. In  case (a), the inner planet is a ``super-Io'', with
intense volcanism, a type of rocky body suggested by Jackson \etal
(2008c). In case (b), the planets could be terrestrial-like bodies:
rocky, with thin atmospheres and modest volcanism.

If the planets are ``mini-Neptunes'', then the orbital history, formation
scenarios, and internal structures are all consistent with previous
models of such bodies. The only constraint tidal evolution can provide is that $i >
15^\circ$ (assuming $Q'_b = 10^5$). As this constraint is less
stringent than the requirements for a terrestrial body, our analysis favors the mini-Neptune model somewhat.

In order to distinguish the super-Earth and mini-Neptunes through
dynamical analyses, the eccentricities need to be determined to within
at least $10^{-3}$. Such high precision is only measurable for planets that undergo secondary transits, but not even a primary transit has (so far) been reported for this
system. Photometric observations of the transit would allow a
calculation of the planet's radius, but this value is not enough to
determine if the planets are super-Earths because of
composition-radius degeneracies exist in current models of small-mass
exoplanets (Adams, Seager \& Elkins-Tanton 2008; Raymond, Barnes \&
Mandell 2008). Therefore dynamical models of the system's history may
be the most effective way to determine the composition of the planets.

We also note that if planet b is gaseous, then it may be undergoing evaporative
mass loss (Baraffe et al. 2004; Hubbard et al. 2007).  Indeed, Raymond
\etal (2008) showed that a 25 $\mearth$ planet at 0.05 AU could be
evaporated to its core in 4-5 Gyr, although this depends on the star's
X-ray history. If planet b had its atmosphere removed in the past,
then the tidal models presented in $\S$ 3 may be inadequate as we have
not considered time-varying $Q'_b$. Furthermore, even if HD 40307 b is
gaseous today, then it may one day be reduced to a solid core, making
it terrestrial.

Our analysis also admits the possibility that HD 40307 b is an exotic
planet, unlike any in our Solar System, with end-member properties
ranging from a volcanic super-Io to mini-Neptunes. $Q'$ values may
range from $\sim 10^3$ to $\sim 10^4$, perhaps resulting from an unusual
internal structure.

\begin{figure}
\plotone{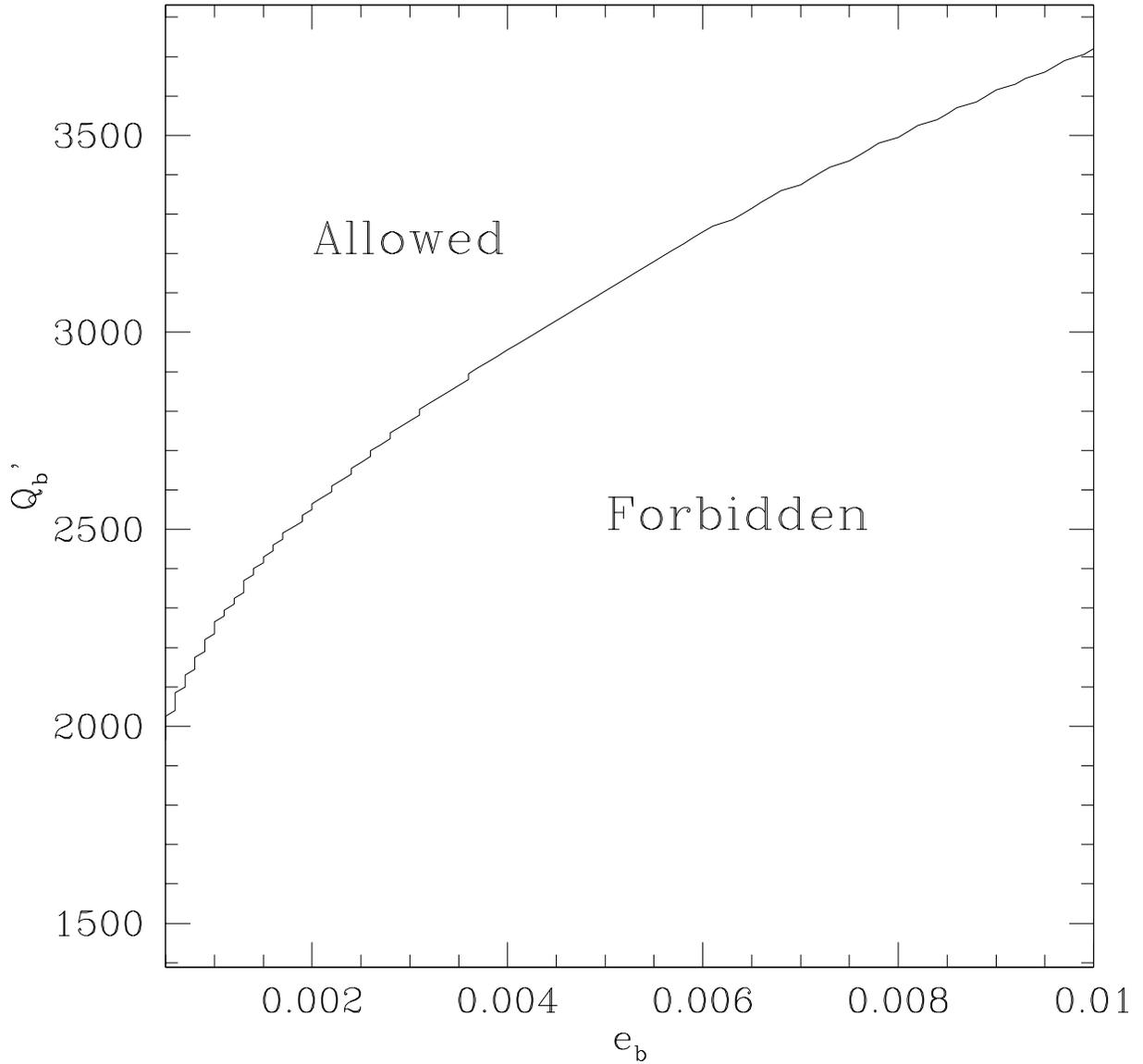}
\figcaption[]{\label{fig:qe} \small{Values of $Q'_b$ and $e_b$ that predict resonance crossing
less than 6.1 Gyr ago (forbidden region) and those that don't (allowed region). This plot demonstrates how the orbital properties of the system can constrain the physical properties of planet b. Note that the Earth's $Q'$ value is $\sim 500$ and Neptune's is $\sim 10^5$.}}
\end{figure}

The dynamical properties of planetary systems
may be used to constrain the physical properties of exoplanets as Fig.\
\ref{fig:qe} demonstrates. As more potentially terrestrial-like planets are detected, dynamical
analyses will continue to play a role in constraining their physical
and orbital properties.

\section{Acknowledgments}
RB, BJ and RG acknowledge support from NASA's PG\&G Program
grant NNG05GH65G. SNR was supported by an appointment to the NASA
Postdoctoral Program at the University of Colorado Astrobiology Center,
administered by Oak Ridge Associated Universities through a contract
with NASA. We also thank an anonymous referee, whose comments greatly
improved this manuscript.

\references
Adams, A.R., Seager, S. \& Elkins-Tanton, L. 2008, ApJ, 673, 1160\\
Adams, F.~C., Laughlin, G., \& Bloch, A.~M.\ 2008, ApJ, 683, 1117\\
Aksnes, K, Franklin, F.A. 2001, AJ, 122, 2734\\
Andrews, S.~M., \& Williams, J.~P.\ 2005, \apj, 631, 1134\\
Banfield, D. \& Murray, N. 1992, Icarus, 99, 390\\
Baraffe, I. \etal 2004, A\&A, 419, L13\\
Barnes, R. \& Quinn, T.R. 2004, ApJ, 611, 494\\
Barnes, R., Raymond, S.N., Jackson, B. \& Greenberg, R. 2008, Astrobiology, 8, 557 \\
Barnes, S.A. 1999, ApJ, 586, 484\\
Bodenheimer, P., Hubickyj, O., \& Lissuaer, J.J. 2000, Icarus, 143, 2\\
Butler, R.P. \etal 2006, ApJ, 646, 505\\
Chiang, E.I., Fischer, D. \& Thommes, E. 2002, ApJ, 564, L105\\
Davies, G. 1999, \textit{Dynamic Earth}, Cambridge UP\\
Deming, D. \etal 2007, ApJ, 667, L199\\
Dickey, J.O. \etal 1994, Science, 265, 482-490\\
Dobrovolskis, A.R. 2007, Icarus, 192, 1\\
Donahue, R.A. 1998, In: Cool Stars, Stellar Systems, and the Sun, Eds: R.A. Donahue \& J.A. Bookbinder, 1235\\
Eggleton, P.P \etal 1998, ApJ, 499, 853\\
Ferraz-Mello, S., Rodr\'iguez, A., \& Hussmann, H. 2008, CeMDA, 101, 171\\
Fortney, J.J., Marley, M.S. \& Barnes, J.W. 2007, ApJ, 659, 1661\\
Gillon, M. \etal 2007, A\&A, 472, L13\\
Goldreich, P. \& Soter, S. 1966, Icarus, 5, 375\\
Gomes, R., Levison, H.F., Tsiganis, K., Morbidelli, A. 2005, Nature, 435, 466\\
Greenberg, R., Barnes, R. \& Jackson, B. 2008, BAAS, 40, 391\\
Hayashi, C.\ 1981, Progress of Theoretical Physics Supplement, 70, 35\\
Hubickyj, O., Bodenheimer, P. \& Lissauer, J.J. 2005, Icarus, 179, 415\\
Hubbard, W.~B., Hattori, M.~F., Burrows, A., Hubeny, I., \& Sudarsky, D.\ 2007, Icarus, 187, 358\\
Hut, P. 1981, A\&A, 99, 126\\
Ikoma, M., Emori, H., \& Nakazawa, K.\ 2001, \apj, 553, 999
Kuchner, M.~J.\ 2004, \apj, 612, 1147\\
Jackson, B., Greenberg, R. \& Barnes, R. 2008a, ApJ, 678, 1396\\
--------. 2008b, ApJ, 681, 1631\\
Jackson, B., Barnes, R. \& Greenberg, R. 2008c, MNRAS, 391, 237\\
Lecoanet, D., Adams, F.C. \& Bloch, A.M. 2008. ApJ, accepted. (astro-ph/08104076)\\
L\'eger, A. \etal 2004, Icarus, 169, L499\\
Lin D.N.C., Bodenheimer P., \& Richardson D.C., 1996, Nature, 380, 606\\
Mamajek, E. E. \& Hillenbrand, L. A. 2008, ApJ, 687, 1284\\
Mardling, R.A 2007, MNRAS, 382, 1768\\
Mardling, R.A. \& Lin, D.N.C. 2002, ApJ, 573, 829\\
Mayor, M \etal 2008, A\&A, submitted (astro-ph/08064587)\\
Noyes, R.W., Weiss, N.O., \& Vaughan, A.H. 1984, ApJ, 287, 679\\
Ossendrijver, M. 2003, A\&AR, 11, 287\\
Parker, E.N. 1993, ApJ, 408, 707\\
Pollack, J.B. \etal 1996, Icarus, 194, 62\\ 
Rafikov, R.~R.\ 2006, \apj, 648, 666\\
Raymond, S.N., Barnes, R. \& Mandell, A.M. 2008, MNRAS, 384, 663\\
Raymond, S.N., Mandell, A.M., \& Sigurdsson, S. 2006, Science, 313, 1413\\
Rivera, E.J. \etal 2005, ApJ, 634, 625\\
Skumanich, A. 1972, ApJ, 171, 565\\
Snellgrove, M.~D., Papaloizou, J.~C.~B., \& Nelson, R.~P.\ 2001, \aap, 374, 1092\\
Soderblom, D.R., Duncan, D.K. \& Johnson, D.R.H. 1991, ApJ, 375, 722\
Thompson, M.J., Christensen-Dalsgaard, J., Miesch, M.S. \& Toomre, J. 2003, A\&RAA, 41, 599\\
Udry, S. \etal 2007, A\&A, 469, L43\\
Weidenschilling, S.~J.\ 1977, \apss, 51, 153\\
Yoder, C.F. 1995, In: \textit{Global Earth Physics. A Handbook of Physical Constants}, edited by T. Ahrens, American Geophysical Union, Washington, D.C.\\
Zhang, K. \& Hamilton, D.P. 2007, Icarus, 188, 386\\
--------. 2008, Icarus, 193, 267\\

\begin{center}
Table 1. Masses and orbits for the HD 40307 planets\\ 
\begin{tabular}{cccccc}
\hline\hline
Planet & $m$ ($\mearth$) & $P$ (d) & $a$ (AU) & $e$ & $T_p$ (JD)\\
\hline
b  & 4.2 & 4.3115 & 0.047 & $0.008 \pm  0.065$ & 2454562.77\\
c  & 6.8 &  9.62 & 0.081 & $0.033 \pm  0.052$ & 2454551.53\\
d  & 9.2 &  20.46 & 0.134 & $0.037 \pm 0.052$ & 2454532.42\\
\end{tabular}
\end{center}
$^a$ Set to zero by Mayor \etal (2008)\\

\clearpage
\clearpage

\clearpage
\clearpage

\clearpage
\clearpage

\clearpage
\clearpage

\section*{Appendix: The Age of HD 40307}
\renewcommand{\theequation}{A\arabic{equation}}
Based on two techniques that use the rotation rate and the magnetic activity
strength respectively, we estimated the age of HD 40307. Dwarf stars with
spectral types ranging from late-F to mid-M have both a radiative and a
convective zone, the interface of which is thought to be responsible for
magnetic field generation and the heating of the upper atmosphere (Parker
1993; Ossendrijver 2003; Thompson et al. 2003), a phenomenon that gives rise
to magnetic activity. These stars begin their lives rotating quickly but
slowly lose angular momentum over time via solar-type winds. This loss of
angular momentum as a function of stellar age has been observed and can be
quantified (e.g. Skumanich 1972). Using the age-rotation relation for a star
with a radiative-convective zone interface, we calculated an age of 6.7$\pm$2.0
Gyr based on a 48 day period, and using the 30\% quoted uncertainty from Barnes (2003; I sequence).

Since the rotation rate is linked to magnetic field generation and the
subsequent chromospheric activity in solar-type stars, a reduced rotation
rate will result in less magnetic activity. Using the cluster derived
age-activity relations from Soderblom Duncan \& Johnson (1991), we calculate an
age (averaging the 3 relations) of 5.4$\pm$1.6 Gyr, based on an $R^{'}_{HK}$ value of -4.99 (Noyes et al.
1984). These results are
confirmed using the age-activity relation of Donahue (1998), which also
yield an age of 5.4 Gyr. Recently, Mamajek \& Hillenbrand (2008) refined the age-activity
relations for F-K dwarfs using additional stellar clusters with
improved age estimates.  Their relations yield an age of 6.3 $\pm$
0.9 Gyr for HD 40307. The Donahue (1998) study uses coeval binary systems
to quantify the uncertainty in these relations. For the early K-type dwarfs,
the age-discrepancy in binary pairs is smaller than the uncertainty in the
age-activity relation, confirming an uncertainty of $\sim$ 1.6 Gyr.

\indent Combining the results from the rotation and activity analysis, we estimate
the age of HD 40307 to be 6.1$\pm$1.6 Gyr.

\end{document}